\documentclass[aps,prl,twocolumn,superscriptaddress,longbibliography]{revtex4-2}

\usepackage{mathrsfs,amssymb,graphics,subfigure,threeparttable}
\usepackage{grap hicx}
\usepackage{amssymb}
\usepackage{enumerate}
\usepackage{amsmath}
\usepackage{fancyhdr}
\usepackage{bm}
\usepackage[squaren]{SIunits}
\usepackage{xcolor}
\usepackage{changes}
\usepackage{soul}
\usepackage{multirow}
\usepackage{hyperref}
\hypersetup{
    colorlinks=true,
    linkcolor=blue,
    filecolor=black,      
    citecolor=blue,
    urlcolor=blue,
    }

\begin{document}



\title{Effects of realistic laser intensity and phase distribution on high-charge laser wakefield acceleration}

\author{Yuhui Xia}
\affiliation{State Key Laboratory of Nuclear Physics and Technology, and Key Laboratory of HEDP of the Ministry of Education, CAPT, Peking University, Beijing 100871, China}
\author{Zhenan Wang}
\affiliation{State Key Laboratory of Nuclear Physics and Technology, and Key Laboratory of HEDP of the Ministry of Education, CAPT, Peking University, Beijing 100871, China}
\author{Ziyao Tang}
\affiliation{State Key Laboratory of Nuclear Physics and Technology, and Key Laboratory of HEDP of the Ministry of Education, CAPT, Peking University, Beijing 100871, China}
\author{Jianghao Hu}
\affiliation{State Key Laboratory of Nuclear Physics and Technology, and Key Laboratory of HEDP of the Ministry of Education, CAPT, Peking University, Beijing 100871, China}
\author{Qianyi Ma}
\affiliation{State Key Laboratory of Nuclear Physics and Technology, and Key Laboratory of HEDP of the Ministry of Education, CAPT, Peking University, Beijing 100871, China}
\author{Yuekai Chen}
\affiliation{State Key Laboratory of Nuclear Physics and Technology, and Key Laboratory of HEDP of the Ministry of Education, CAPT, Peking University, Beijing 100871, China}
\author{Letian Liu}
\affiliation{State Key Laboratory of Nuclear Physics and Technology, and Key Laboratory of HEDP of the Ministry of Education, CAPT, Peking University, Beijing 100871, China}
\author{Zhiyan Yang}
\affiliation{State Key Laboratory of Nuclear Physics and Technology, and Key Laboratory of HEDP of the Ministry of Education, CAPT, Peking University, Beijing 100871, China}
\author{Hui Zhang}
\affiliation{State Key Laboratory of Nuclear Physics and Technology, and Key Laboratory of HEDP of the Ministry of Education, CAPT, Peking University, Beijing 100871, China}
\author{Chenxu Wang}
\affiliation{State Key Laboratory of Nuclear Physics and Technology, and Key Laboratory of HEDP of the Ministry of Education, CAPT, Peking University, Beijing 100871, China}
\author{Haoyang Lan}
\affiliation{State Key Laboratory of Nuclear Physics and Technology, and Key Laboratory of HEDP of the Ministry of Education, CAPT, Peking University, Beijing 100871, China}
\affiliation{Beijing Laser Acceleration Innovation Center, Beijing 100871, China}
\author{Di Wu}
\affiliation{State Key Laboratory of Nuclear Physics and Technology, and Key Laboratory of HEDP of the Ministry of Education, CAPT, Peking University, Beijing 100871, China}
\affiliation{Beijing Laser Acceleration Innovation Center, Beijing 100871, China}
\author{Xiuhong Yang}
\affiliation{State Key Laboratory of Nuclear Physics and Technology, and Key Laboratory of HEDP of the Ministry of Education, CAPT, Peking University, Beijing 100871, China}
\affiliation{Beijing Laser Acceleration Innovation Center, Beijing 100871, China}
\author{Yixing Geng}
\affiliation{State Key Laboratory of Nuclear Physics and Technology, and Key Laboratory of HEDP of the Ministry of Education, CAPT, Peking University, Beijing 100871, China}
\affiliation{Beijing Laser Acceleration Innovation Center, Beijing 100871, China}
\author{Yanying Zhao}
\email{zhaoyanying@pku.edu.cn}
\affiliation{State Key Laboratory of Nuclear Physics and Technology, and Key Laboratory of HEDP of the Ministry of Education, CAPT, Peking University, Beijing 100871, China}
\affiliation{Beijing Laser Acceleration Innovation Center, Beijing 100871, China}
\author{Xueqin Yan}
\affiliation{State Key Laboratory of Nuclear Physics and Technology, and Key Laboratory of HEDP of the Ministry of Education, CAPT, Peking University, Beijing 100871, China}
\affiliation{Beijing Laser Acceleration Innovation Center, Beijing 100871, China}
\affiliation{Institute of Guangdong Laser Plasma Technology, Baiyun, Guangzhou, 510540, China}
\author{Xinlu Xu}
\email{xuxinlu@pku.edu.cn}
\affiliation{State Key Laboratory of Nuclear Physics and Technology, and Key Laboratory of HEDP of the Ministry of Education, CAPT, Peking University, Beijing 100871, China}
\affiliation{Beijing Laser Acceleration Innovation Center, Beijing 100871, China}

\date{\today}

\begin{abstract}
Laser wakefield acceleration (LWFA) can produce relativistic electron beams and various secondary particles in centimeter-long plasmas, making it a valuable particle source with important applications in many disciplines. In this work, we examine the effects of non-ideal transverse intensity and phase distribution of laser pulses on LWFA  through both experimental measurements and particle-in-cell simulations. The complex transverse profile of the 75 TW laser pulses reduces the self-focused intensity in plasma compared with a transversely Gaussian laser. Furthermore, the sheath structure of the nonlinear plasma wake excited by realistic laser pulses is wider and more complicated than that of a Gaussian laser. These hinder the injection of the plasma electrons. As the laser pulse propagates through the plasma, its intensity profile gradually becomes elliptical and drives a plasma wake with a sharp sheath near the azimuths of the major axis, leading to an injection. When using a realistic laser profile in simulations, both the charge and energy of injected electrons closely match experimental results {($\sim200$ pC of charge and $\sim 200$ MeV peak energy)}, whereas the Gaussian laser simulations produce much higher charge ($\sim500$ pC). Our findings reveal the difference in injection dynamics between LWFA driven by non-ideal laser pulses and those driven by Gaussian pulses, and are useful for applications of LWFA which demand high-charge electron beams. 
\end{abstract}


\maketitle


\section{\label{sec:level1}I. Introduction}
{Laser wakefield acceleration (LWFA) \cite{PhysRevLett.43.267} can generate monoenergetic electron beams with hundreds of MeV to $\sim10$ GeV energies \cite{faure2004laser,
geddes2004high, mangles2004monoenergetic, leemans2006gev, PhysRevLett.122.084801, aniculaesei2024acceleration, picksley2024matched}, $\sim$femtosecond durations \cite{lundh2011few, buck2011real, PhysRevLett.125.014801, huang2024electro, laberge2024revealing, ma2025rec}, and $\sim$mrad divergences \cite{leemans2014multi, ma2015multiple, wang2016high} in centimeter-long plasmas \cite{RevModPhys.81.1229, joshi2020perspectives}.} As compact and economical sources of relativistic electrons, LWFA holds significant potential for applications in fields ranging from next-generation colliders to compact x-ray free-electron lasers \cite{wang2021free, labat2023seeded,vh62-gz1p}. Meanwhile, ultrafast secondary x-ray and $\gamma$-ray radiation produced through betatron radiation \cite{RevModPhys.85.1}, Compton scattering \cite{PhysRevLett.96.014802} and bremsstrahlung based on LWFA have been applied in x-ray phase contrast imaging \cite{fourmaux2011single}, microcomputed tomography \cite{cole2018high}, x-ray absorption spectroscopy \cite{PhysRevLett.123.254801}, and photonuclear reactions \cite{wu2024ultra}, and exhibited advantages in terms of compactness, collimation, ultrashort duration, and brightness. These applications, which benefit from a large number of photons, usually demand a high charge of relativistic electrons while having loose requirements on beam quality (energy spread and emittance). Various schemes have been explored experimentally to improve the injected charge. Several hundreds of $\pico\coulomb$ monoenergetic electron beams with $\sim 100~\mega\electronvolt$ energy have been generated using widespread 100-$\tera\watt$, joule-level laser systems through self-injection \cite{mcguffey2012experimental, li2017generation}, ionization injection \cite{couperus2017demonstration}, density downramp injection \cite{PhysRevX.10.041015, liu2023generation} and nanoparticle-insertion scheme \cite{xu2022nanoparticle}.

In most studies on LWFA, the laser pulse driver is assumed to have a diffraction-limit Gaussian transverse profile. However, realistic laser pulses delivered by state-of-the-art ultrafast high-power laser systems often deviate from the ideal Gaussian beam in both intensity and phase due to complex amplification and compression procedures \cite{pariente2016space, alonso2024space, howard2025single}. During the rapid development of LWFA over the past decade, researchers have increasingly focused on the fine dynamics of the evolution of laser pulses in plasmas, the injection and acceleration of electrons, and the reproducibility of LWFA \cite{maier2020decoding, PhysRevAccelBeams.26.032801, Amodio_2025, j5md-sckl}. Consequently, the imperfections of realistic laser pulses and their impact on LWFA have become increasingly significant. Previous studies have shown that a laser pulse with an imperfect wavefront \cite{mangles2009controlling, beaurepaire2015effect} or intensity distribution \cite{glinec2008direct, corde2013observation, nakanii2016effect} drives an inhomogeneous transverse wakefield, which affects the transverse distribution of the electrons. Additionally, non-Gaussian lasers have been found to strongly degrade the betatron x-ray emission \cite{ferri2016effect}. Recent works have demonstrated that a specific complex wavefront improves the performance of LWFA \cite{he2015coherent, oumbarek2023notable}.

In this paper, we synthetically use experiments and particle-in-cell (PIC) simulations to demonstrate that the imperfection of a realistic laser pulse substantially reduces the injected electron charge. A 35 $\femto\second$ Ti:Sapphire laser pulse with a 1.3 J energy is incident on a helium gas jet and {produces electrons with a peak energy of $\sim200$ MeV and a total charge of $\sim$200 pC}. The charge is significantly lower than PIC simulation results using a transversely Gaussian laser with the same peak power. By importing the realistic laser distribution reconstructed using the Gerchberg–Saxton (GS) algorithm into the PIC simulation, we find that the non-ideal intensity and phase distribution of the laser lower the self-focused laser intensity, blur the sheath structure of the nonlinear plasma wake and hinder the injection at the initial stage of the plasma. As the relativistic laser pulse propagates through the plasma, its intensity profile evolves into an ellipse and enables the injection. The charge and energy of the injected electrons in simulations using {realistic} laser pulse agree well with experiments. In Sec II, we present the experimental results of the LWFA. In Sec III, a comparison is performed between PIC simulations of LWFA driven by a Gaussian laser pulse, an elliptical laser pulse and a reconstructed realistic laser pulse, and the differences in laser pulse evolution, plasma wake excitation and electron injection are discussed. A summary of our finding is given in Sec. IV. 

\section{\label{sec:level2}II. Experimental setup and results}
\begin{figure}
\centering
\includegraphics[width=1.0\linewidth]{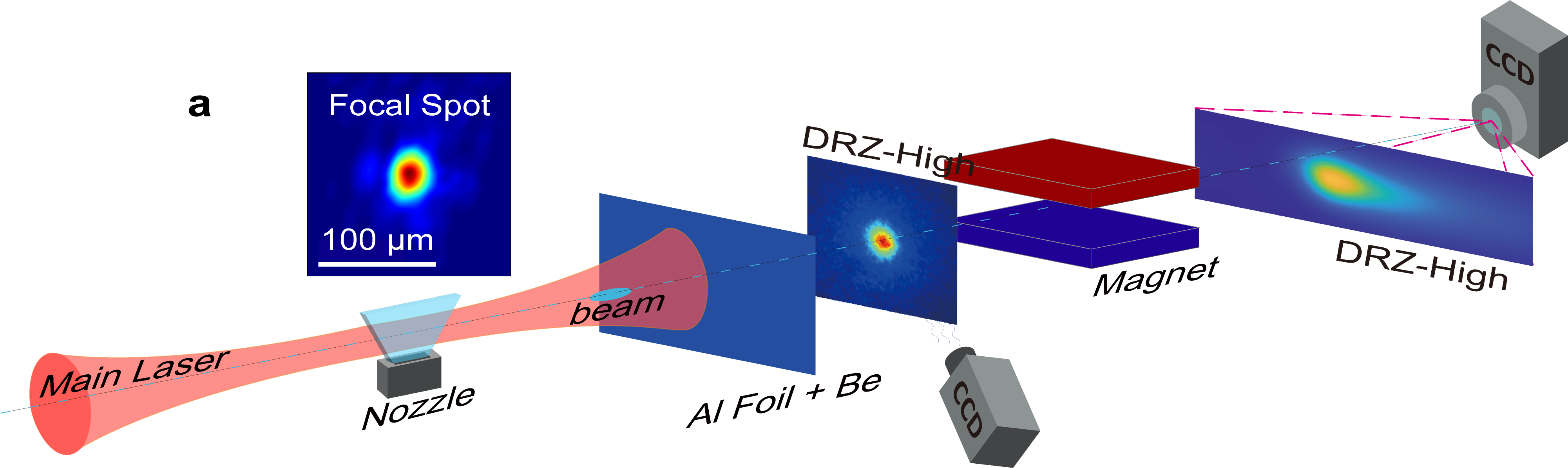}
\includegraphics[width=0.5\linewidth]{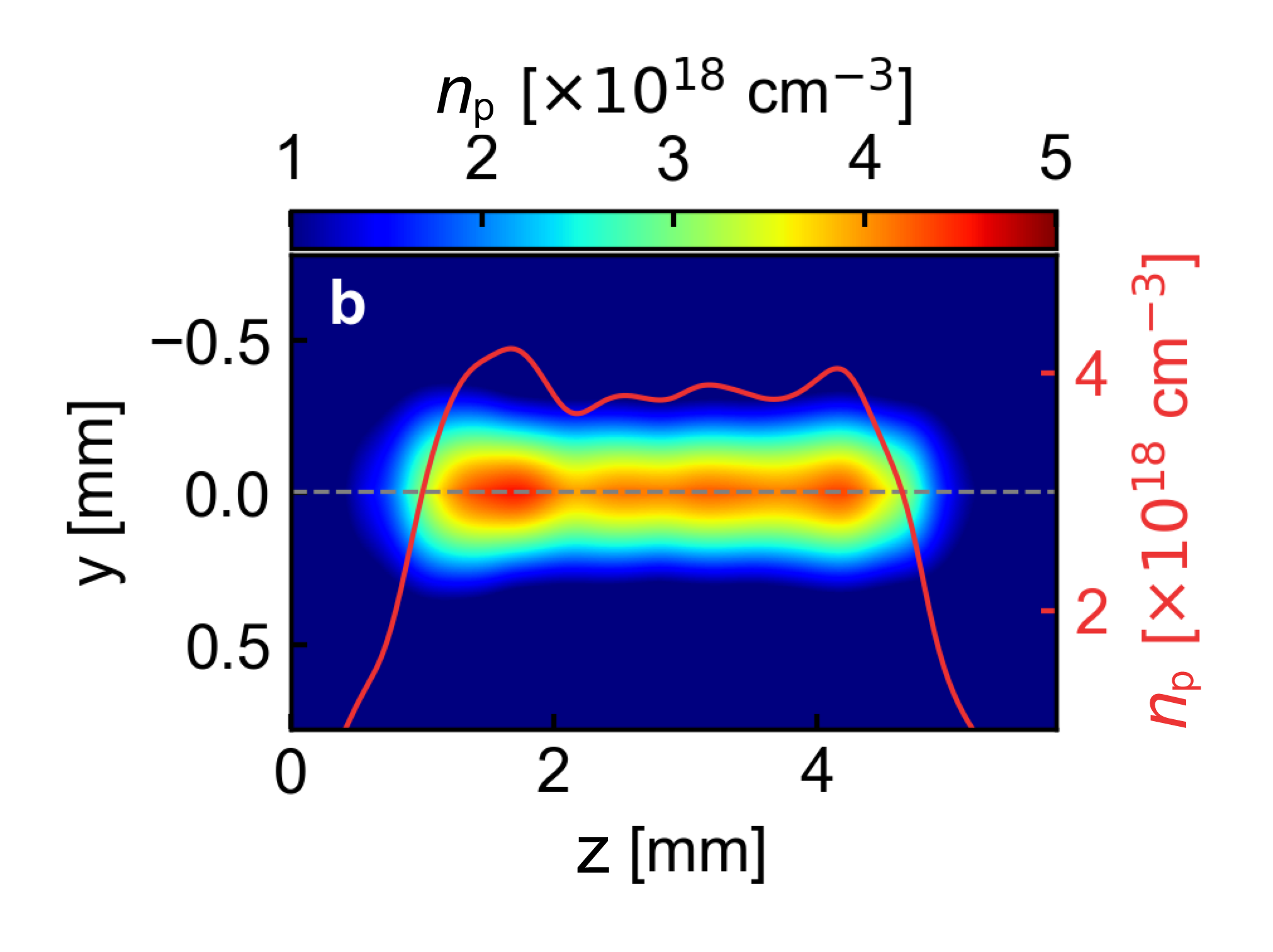}\includegraphics[width=0.5\linewidth]{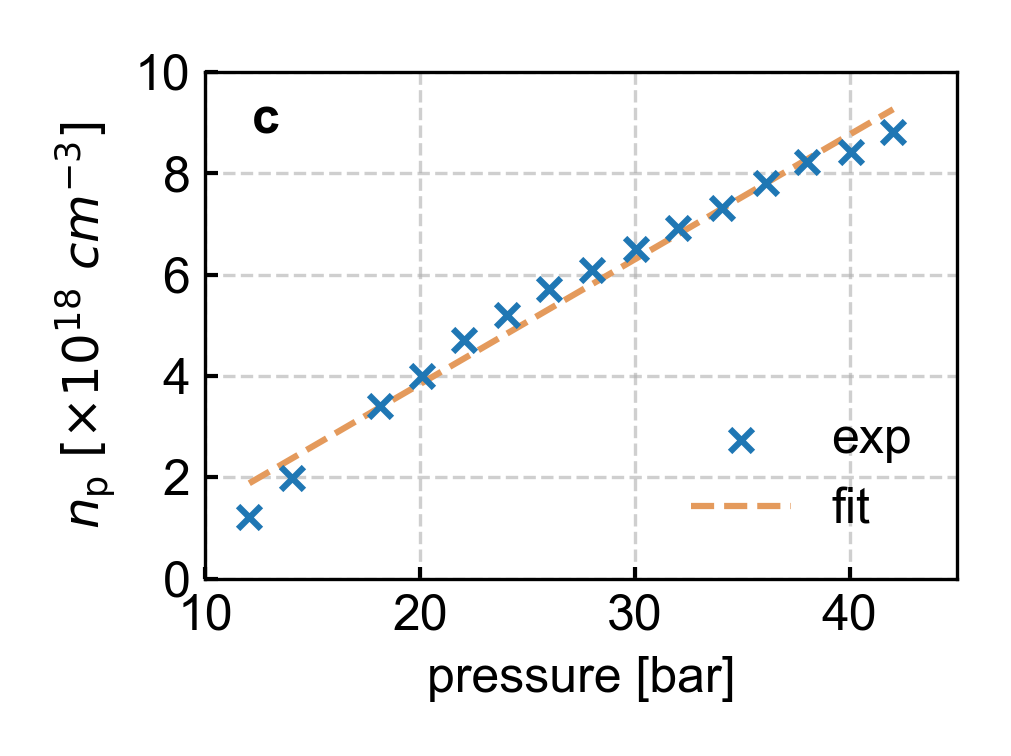}
\caption{\label{fig:scheme} (a) Schematic of the experimental setup (not on scale). The inset is a typical spatial intensity profile of the laser pulse at the focus. A 10 $\micro\meter$ thick aluminum film is placed near the gas nozzle to block the laser pulse while allowing relativistic electrons to pass through it with minimal impact. (b) A typical plasma electron density distribution measured by the Michelson interferometer and the axial plasma density profile (red line). (c) The dependence of the plasma density on the backing pressure. The plasma density is measured at a height of 2 mm above the nozzle.}
\end{figure}

A schematic of the experiment is shown in Fig. \ref{fig:scheme}(a). The experiment was carried out at the Compact Laser Plasma Accelerator Laboratory (CLAPA) at Peking University, utilizing a 5 $\hertz$ high power Ti:sapphire laser system. A $x$-direction (horizontally) linearly polarized laser pulse had a full width at half maximum (FWHM) pulse duration of $35~\femto\second$ and was focused down to a spot size of $w_0 \approx 25.9~\micro\meter$ via an F/12.5 off-axis parabolic mirror. A typical spatial intensity distribution of the laser focus spot is shown in the inset of Fig. \ref{fig:scheme}(a). The laser energy within the $1/e^2$ radius was 1.3 J, accounting for 51\% of the total energy. This corresponds to a peak normalized vector potential $a_0\equiv \frac{eE_0}{mc\omega_0} \approx 1.44$, where $E_0$ and $\omega_0$ are the electric field and the frequency of the laser, $m$ and $e$ are the electron mass and charge, and $c$ is the speed of light in vacuum. A 4-$\milli\meter$-long and 1-$\milli\meter$-wide De Laval supersonic gas nozzle provided a helium gas target. A probe laser beam [not shown in Fig. \ref{fig:scheme}(a)], perpendicular to the main laser propagation axis and containing $2\%$ energy of the main pulse, was used to diagnose the plasma density in a Michelson interferometer. A typical plasma electron density distribution at a height of 2 mm above the nozzle when the backing pressure was 21 bar is presented in Fig. \ref{fig:scheme}(b), and the red line shows the on-axis density profile. The plasma density had an approximately linear dependence on the backing pressure as shown in Fig. \ref{fig:scheme}(c). 

A 20 $\centi\meter\times$23 $\centi\meter$ dipole magnet with 0.8 $\tesla$ peak magnetic field was placed 80 $\centi\meter$ away from the gas nozzle to measure the beam energy spectrum with a reliable energy range of $70~\mega\electronvolt$ to $350~\mega\electronvolt$. Two GOS ($\rm Gd_2O_2S:Tb$, $\rm \rho_{GOS}=7.44$ $\rm g/cm^{3}$) scintillators (DRZ-High, phosphor surface loading of $145$ $\rm mg/cm^2$) were individually positioned at the entrance and the exit of the magnet. Two imaging systems with Andor CCD recorded the divergence and the pointing angle of the electron beams at the first screen and the energy spectrum at the second screen. The beam charge was calculated using the measured signals at the second screen by taking account of the photon generation and transmission efficiencies from the scintillator to the CCD. {The scintillator was calibrated with a conventional electron accelerator.}

Pure helium gas was used and we relied on the evolution of the laser pulse driver to trap electrons \cite{tsung2004near, mangles2004monoenergetic, geddes2004high, faure2004laser, kalmykov2009electron}. For the interaction between an ultrashort and ultra-intense laser pulse and an underdense plasma, self-focusing, self-phase modulation and pump depleting occur \cite{mori1997physics}, and the laser pulse and the excited plasma wake evolve significantly. When the laser peak intensity is high and the wake expands (i.e., the phase velocity of the wake tail is significantly less than the speed of light), some sheath electrons of the nonlinear plasma wake with high forward velocities can be trapped and accelerated to hundreds of MeV energy. In the experiments, the relative position between the laser focal plane and the gas nozzle and the plasma density were examined to optimize the charge of the injected beams. We moved the gas nozzle along the laser prorogation path to scan the relative focal position from $z_\mathrm{f}=-2.4~\milli\meter$ to $0.4~\milli\meter$ and found that the injected charge doesn't exhibit a strong dependence on $z_\mathrm{f}$. Here $z_\mathrm{f}=0~\milli\meter$ indicates the laser is focused at the middle of the gas nozzle. This may be due to the long vacuum Rayleigh length of the laser we used, $z_\mathrm{R}\approx 2.6~\milli\meter$. We fixed the relative focal position as $z_\mathrm{f}=-1~\milli\meter$ in the subsequent experiments.

\begin{figure}
\centering
\includegraphics[width=1\linewidth]{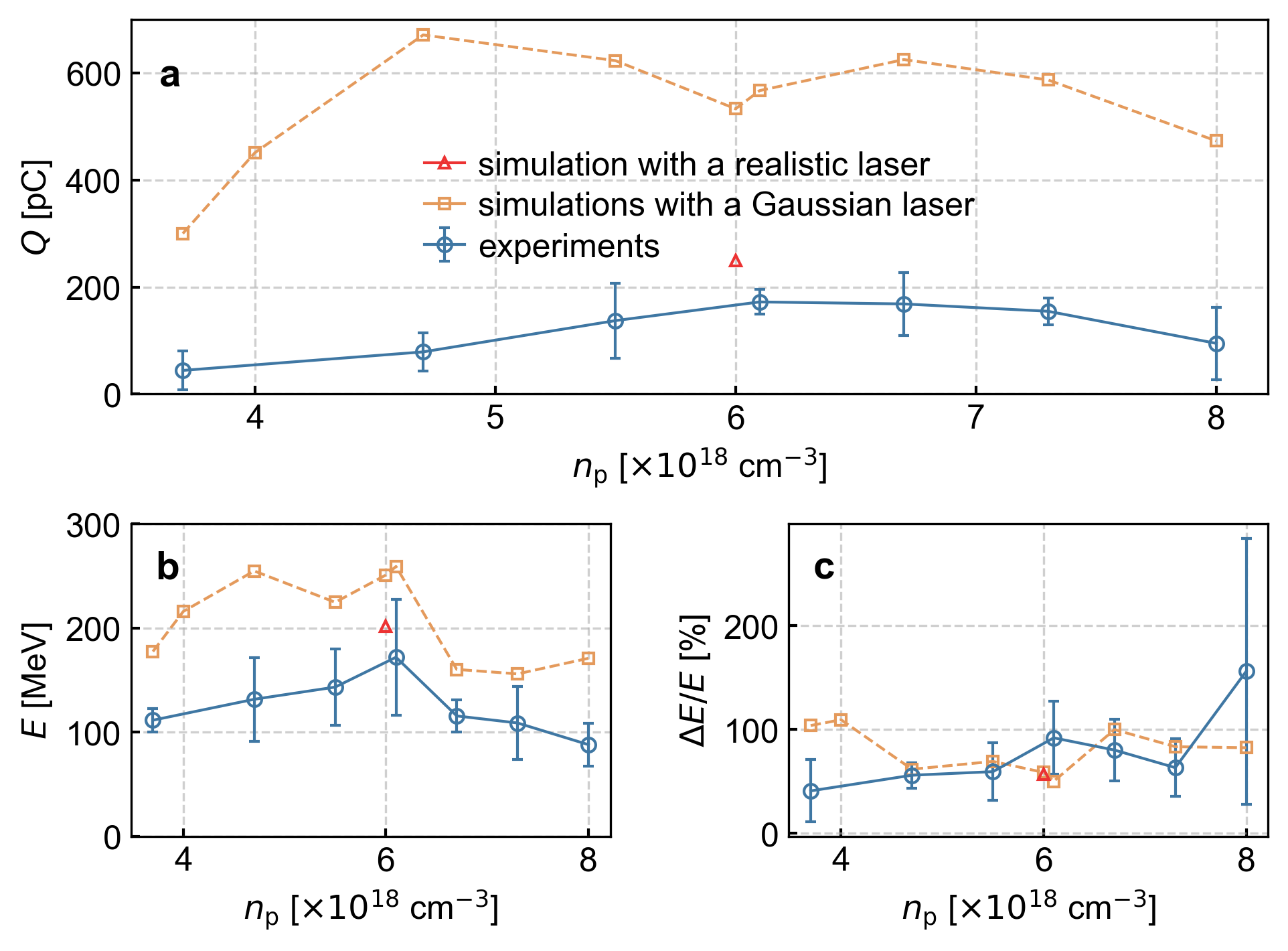}
\caption{\label{fig:exp_results} Experimental results of the injected electron beams when scanning the plasma density $n_\mathrm{p}$. The dependence of the injected charge (a), the peak energy (b) and the relative FWHM energy spread on $n_\mathrm{p}$. Note $z_\mathrm{f}$ is fixed as $-1~\milli\meter$ in these experiments. The presented results are averaged over 5-15 shots at each $n_\mathrm{p}$. The orange squares show the PIC simulation results when using a transversely Gaussian laser pulse driver and the red triangle is the result with a reconstructed laser pulse.}
\end{figure}

The plasma density $n_\mathrm{p}$ plays a critical role in the laser pulse evolution and the injection of electrons. We scanned the plasma density by changing the backing pressure of the nozzle. The measured beam charge under different $n_\mathrm{p}$ is depicted in Fig. \ref{fig:exp_results}(a). When $n_\mathrm{p}<3\times10^{18}~\centi\meter^{-3}$, the self-focusing of the laser pulse is weak, and its peak $a_0$ is less than 3.5 as revealed in PIC simulations using transversely Gaussian laser pulses, resulting in no detectable experimental signal in the screen. The injected charge increased from $44~\pico\coulomb$ to $172~\pico\coulomb$ as the density increased from $3.7\times10^{18}~\centi\meter^{-3}$ to $6.1\times10^{18}~\centi\meter^{-3}$. The injected charge reached its maximum when $n_\mathrm{p}=6.1\times10^{18} ~\centi\meter^{-3}$ and then decreased as the density continued to increase. For high density plasmas, the injected electrons are decelerated at the tail of the plasma due to the strong pump depletion of the laser pulse and the dephasing \cite{liu2023generation}, resulting a reduction in the measured charge which only counts electrons with $\geq 70$ MeV energy. 

The peak energy of the electron beams are shown in Fig. \ref{fig:exp_results}(b). As $n_\mathrm{p}$ increased from $3.7\times10^{18}~\centi\meter^{-3}$ to $6.1\times10^{18}~\centi\meter^{-3}$, the beam energy also increased. This is probably caused by an increasing axial plasma wakefield ($E_z\propto \sqrt{n_\mathrm{p}}$) and fixed plasma length. After $n_\mathrm{p}=6.1\times10^{18}~\centi\meter^{-3}$, the energy decreased as $n_\mathrm{p}$ increased, which is due to the dephasing between the electrons and the laser driver \cite{PhysRevSTAB.10.061301}. The relative FWHM (full width at half maximum) energy spread of the beams was large, $\sim 100\%$ as shown in Fig. \ref{fig:exp_results}(c). 

PIC simulations employing laser pulses with transversely Gaussian profiles reveal electron injection with significantly higher charge than observed in experimental measurements [orange squares in Fig. \ref{fig:exp_results}(a)]. For example, the injected charge when $n_\mathrm{p}=6\times 10^{18}~\centi\meter^{-3}$ is 533 pC in simulations and 172 pC in experiments. This discrepancy may be caused by the imperfection of the transverse distribution of the realistic laser.

\section{III. Effects of realistic laser intensity and phase distribution on injection}

In the experiments, we measured the intensity profiles of the laser pulse at two planes along the propagation axis and retrieved the phase distribution utilizing the GS algorithm. {Note that this phase retrieved algorithm treats the laser pulse as an 800 nm monochromatic carrier and thus cannot resolve the wavelength-dependent information.} Details on how to get the three-dimensional (3D) field distribution of the laser pulse can be found in the Appendix. We then imported the realistic field distribution of the laser into quasi-three-dimensional (Q3D) version \cite{davidson2015implementation} of the particle-in-cell (PIC) code OSIRIS \cite{fonseca2002high} to study its driven LWFA. The simulation used a moving window with a box size of $706.8k_\mathrm{0}^{-1}$ and $684k_\mathrm{0}^{-1}$ and $3534 \times 684$ cells along the $z$ and $r$ directions, respectively, where $k_0=\frac{2\pi}{0.8}~\micro\meter^{-1}$ is the laser wavenumber. The grid sizes were $\Delta z = \frac{1}{5}k_\mathrm{0}^{-1}~(0.025~\micro\meter)$ and $\Delta r = k_\mathrm{0}^{-1}~( 0.127~\micro\meter) < 0.44~\micro\meter$ (the spatial resolution of the laser spot diagnostic system). The time step $\Delta t = 0.1\omega_\mathrm{0}^{-1}$ satisfied the Courant-Friedrichs-Lewy condition. Azimuthal modes from $m=0$ to $6$ were used and 48 macro-particles per cell were initialized to represent the plasma electrons for the realistic laser case. The realistic laser pulse (labeled as `case r') had $1/e^2$ radii of $w_{x} = 22.4~\micro\meter$ (polarized direction) and $w_{y} = 29.3~\micro\meter$ (orthogonal direction), and was focused at 1.2 mm downstream ($z_\mathrm{f}=-1~\milli\meter$) from the plasma entrance with a peak $a_\mathrm{0}=1.44$. If we fit the intensity profile at the focus with an ellipse, its major axis was inclined at $65.7^\circ$ to the $x$-axis. The Zernike coefficients of the realistic laser is shown in Fig. \ref{figure:fig8} in the Appendix.

\begin{figure}
\centering
\includegraphics[width=1.0\linewidth]{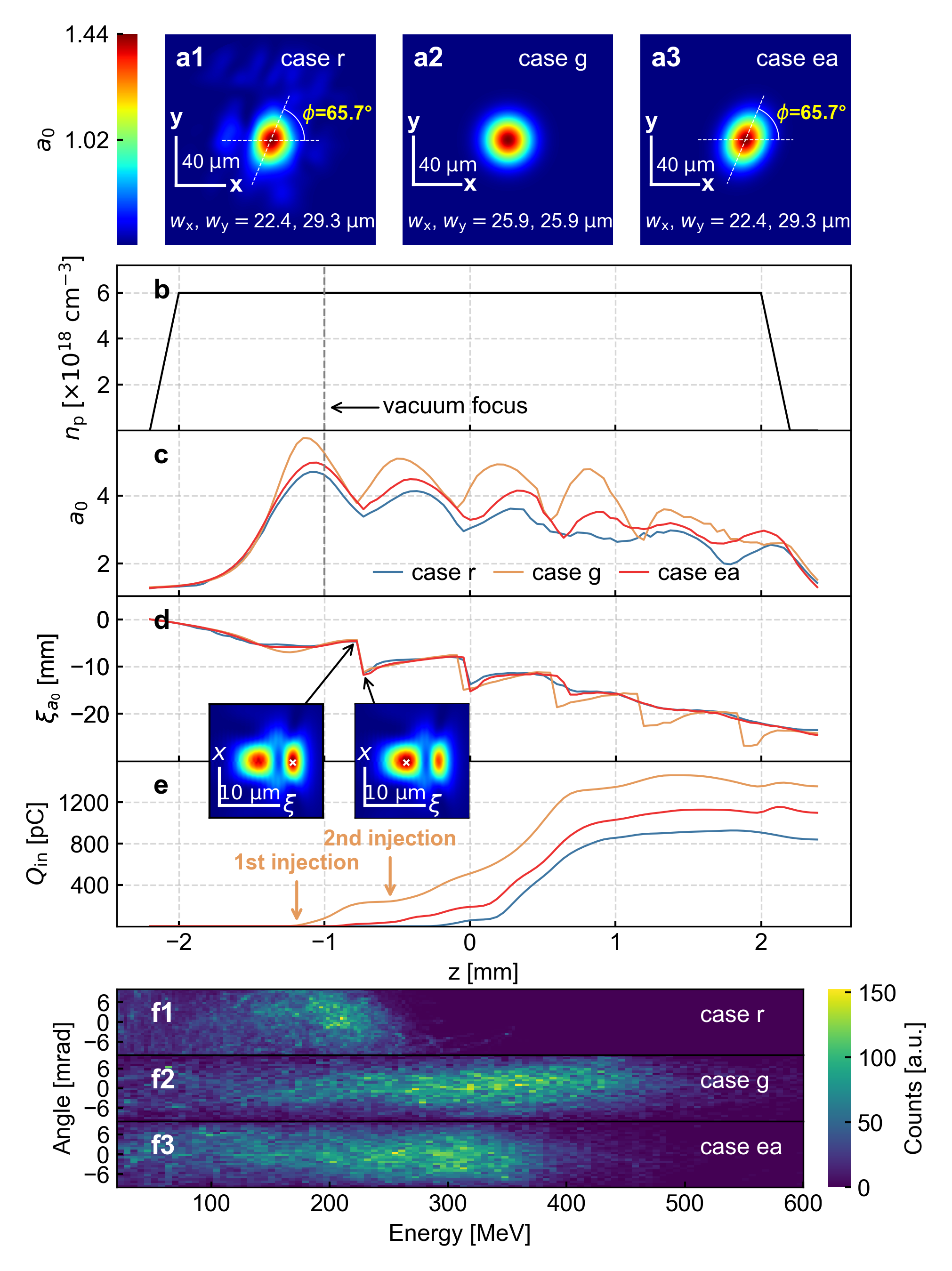}
\caption{\label{fig:laser_para_evo} (a1)-(a3) The transverse intensity distribution of the lasers at the vacuum focus. (b) The plasma density profile. The black dashed line indicates the vacuum focal plane. The evolution of $a_0$ (c) and their axial locations $\xi_{a_0}$ (d) for three cases. The two insets shows the $\xi-x$ intensity profile of the Gaussian laser at different propagation distances, with white crosses marking the locations of the intensity peaks. (e) The evolution of the injected charge for three cases. (f1)-(f3) The energy spectra of the injected beam at the plasma exit for three cases.}
\end{figure}

For comparison purposes, an additional Q3D simulation with 2 azimuthal modes ($m=0, 1$) was performed using an ideal Gaussian laser pulse (labeled as `case g'). The spot size of the Gaussian laser was $w_{x}=w_{y}=25.9~\micro\meter$. {To isolate the physical processes related to an elliptical laser intensity distribution, we also performed one more simulation with 4 azimuthal modes using a laser pulse whose intensity profile was derived from the elliptical fit of the realistic laser (labeled as `case ea').} It also had the same astigmatism terms as the realistic laser, i.e. the Zernike coefficients $Z_{2}^{-2}=0.745~\radian$ and $Z_{2}^{2}=0.656~\radian$, resulting in a 1.09 mm separation between the tangential focus and the sagittal focus. This allowed us to specifically investigate the effect of nonaxisymmetric features of the laser driver on the electron injection in LWFA. Fig. \ref{fig:laser_para_evo}(a1)-(a3) shows the peak intensity profiles of these three lasers at the vacuum focus. The lasers have the same peak $a_\mathrm{0}=1.44$ at the vacuum focus and the same Gaussian temporal envelop with a FWHM duration of 35 fs.

As shown by the black line in Fig. \ref{fig:laser_para_evo}(b), a pre-ionized plasma with $n_\mathrm{p}=6\times10^{18}~\centi\meter^{-3}$, featuring a 200-$\mathrm{\micro\meter}$-long linear upramp, a 4-mm-long plateau and a 200-$\mathrm{\micro\meter}$-long linear downramp, was used in the simulations. Fig. \ref{fig:laser_para_evo}(c) shows the evolution of the peak $a_\mathrm{0}$ for three cases. Since the laser peak power is $\sim75$ TW (on target), much higher than the self-focusing critical power ($P_\mathrm{c}\approx 17 \frac{\omega_0^2}{\omega_\mathrm{p}^2}~\giga\watt\approx 4.9~\tera\watt$ for $n_\mathrm{p}=6\times10^{18}~\centi\meter^{3}$ and 800 nm laser wavelength) \cite{sun1987self, sprangle1987relativistic}, the laser pulse is self-focused by the plasma. The peak $a_0$ of the laser increases as the laser propagates inside the plasma and reaches its maximum value at $z\approx -1.1~\milli\meter$ (before the vacuum focus $z_\mathrm{f}=-1~\milli\meter$). The maximum $a_0$ for the Gaussian laser case is 5.7, significantly higher than the vacuum focused value ($a_0=1.44$). The elliptical laser and realistic laser cases exhibit lower maximum $a_0$ values of 5.0 and 4.7, respectively. As the laser pulse propagates further inside the plasma, the mutual interaction of the self-focusing and the diffraction results in a oscillation of $a_0$ with a decreasing amplitude. The oscillation period is $\sim 0.64~\milli\meter$.  

The temporal envelop of the laser evolves under rich laser-plasma interactions, e.g., the self-steeping, the self phase modulation and the group velocity dispersion \cite{mori1997physics}. As shown in the left inset in Fig. \ref{fig:laser_para_evo}(d), the initial Gaussian temporal envelope with a duration of 35 fs gradually splits into two pulses at $z \approx -0.9~\milli\meter$ \cite{tsung2004near}. At the moment of split, the $a_0$ of the preceding pulse is higher than that of the succeeding pulse. As the preceding pulse depletes its energy into the plasma and diffracts, the intensity of the succeeding pulse gradually surpasses it [the right inset of Fig. \ref{fig:laser_para_evo}(d)]. This causes a step-like jump of the axial position where the peak $a_0$ is located ($\xi_{a_0}$, indicated by the white crosses in the insets), as shown in Fig. \ref{fig:laser_para_evo}(d). Here $\xi\equiv ct - z$ is introduced to express the axial location in a frame moving the speed of light. When the energy of the preceding pulse is nearly exhausted, the succeeding pulse repeats the split and causes the second jump of $\xi_{a_0}$ at $z\approx 0~\milli\meter$. This process repeats more times for the Gaussian laser case than the other cases. 

Figure \ref{fig:laser_para_evo}(e) depicts different injection dynamics for three cases. In case g, the first injection occurs in $-1.2~\milli\meter<z<-0.8~\milli\meter$ where $a_0$ reaches its peak ($a_0=5.7$). Electrons with $\sim200$ pC charge are injected and form a short beam with a peak current of $\sim30$ kA. The second injection occurs at $z\approx -0.4~\milli\meter$ and continues until $z\approx 0.7~\milli\meter$. In contrast, there is no injection when the $a_0$ reaches its maximum ($z\approx -1.1~\milli\meter$) for the elliptical laser and realistic laser cases. There is an injection of 184 pC charge between $z\approx-0.9~\milli\meter$ to $z\approx0~\milli\meter$ for the elliptical laser case while the injected charge is only 49 pC for the realistic laser case. A significant injection happens between $z\approx0.1~\milli\meter$ and $z\approx0.9~\milli\meter$ for these two cases. At the plasma exit, the injected charge is 1.35 nC for the Gaussian laser case, 1.1 nC for the elliptical laser case, and 0.84 nC for the realistic laser case. To make a comparison with the experiment results which only counted electrons with energy above 70 MeV and divergence below 10 mrad, we selected electrons from simulations with the same criterion and the results were 533 pC in case g, 417 pC in case ea and 249 pC in case r. Recall that the experimental result at this plasma density and the vacuum focal position was $172\pm23$ pC, which agrees with the simulation using a realistic laser pulse. Figures \ref{fig:laser_para_evo}(f1)-(f3) show the electron energy spectra at the plasma exit. {The peak energies are 251, 232 and 202 MeV for case g, ea and r, respectively}, which all agree reasonable with the experimental result (172$\pm56$ MeV).

\begin{figure}
\centering
\includegraphics[width=1.0\linewidth]{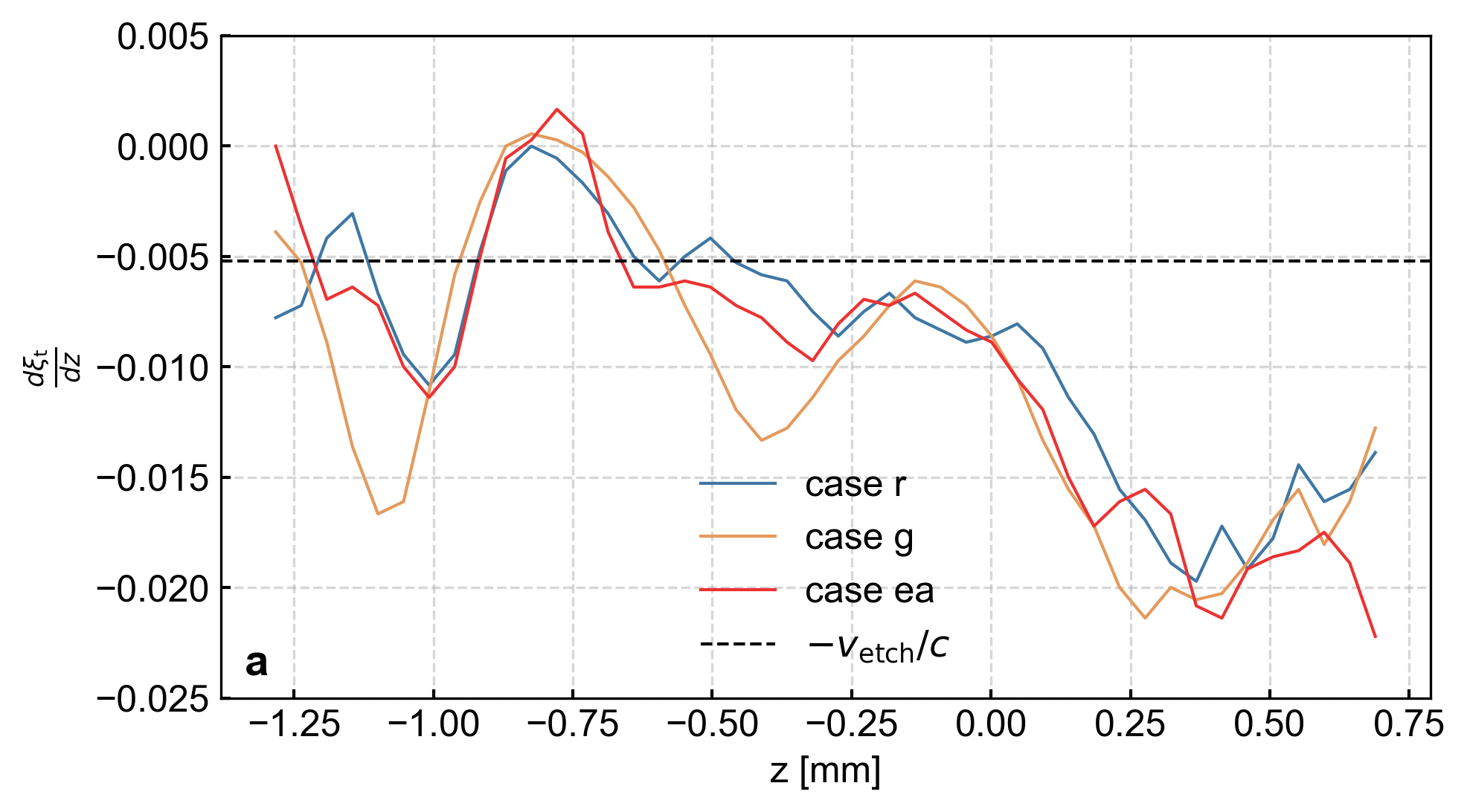}
\includegraphics[width=1.0\linewidth]{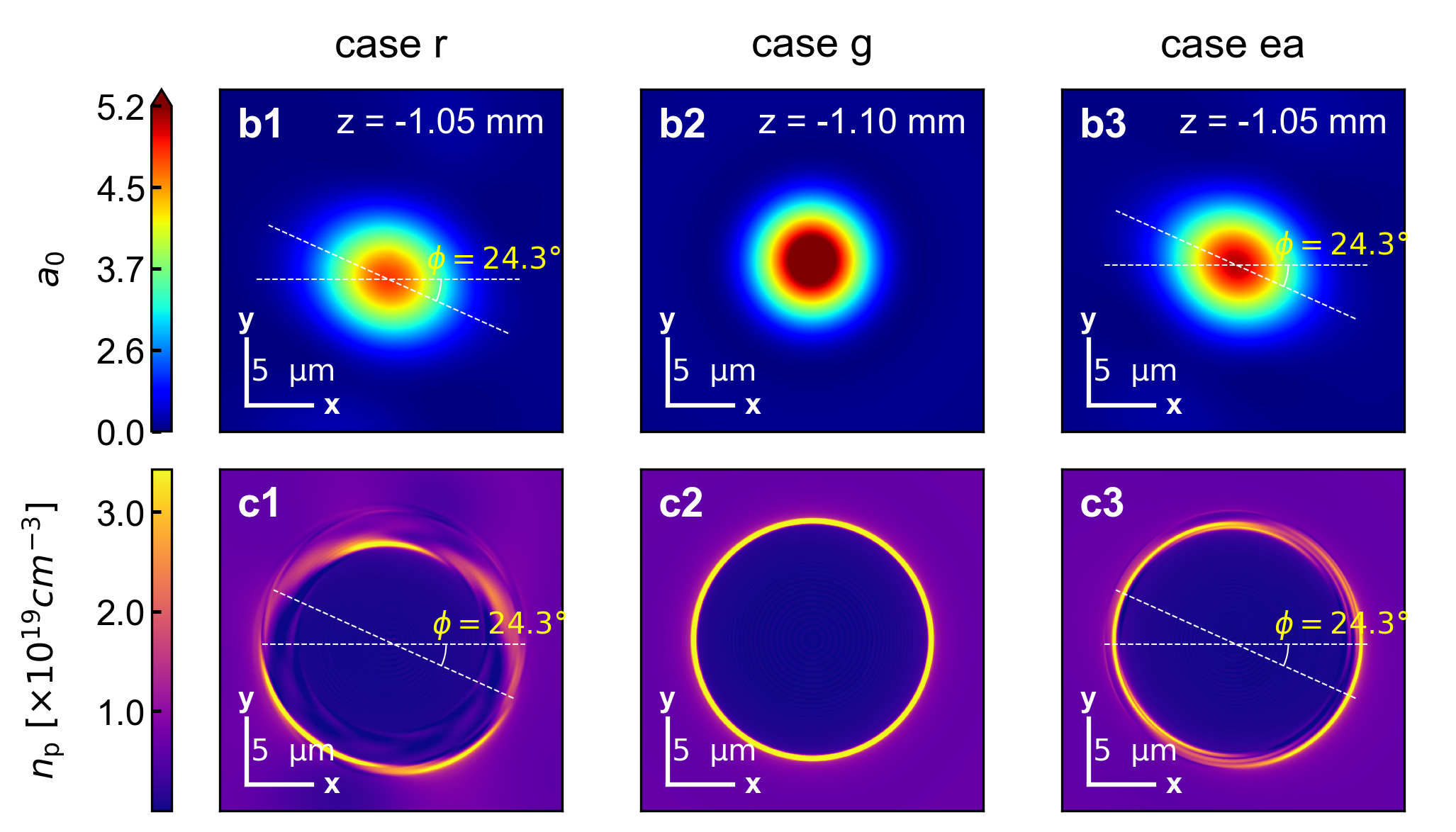}
\caption{\label{fig:wake_velocity} (a) The evolution of $\frac{\mathrm{d}\xi_\mathrm{t}}{\mathrm{d}z}$ for three cases. The black dashed lines is the theoretical value of $-\frac{v_\mathrm{etch}}{c}$. (b1)-(b3) The transverse intensity profile of the $\xi_{a_0}$-slice of the lasers at the propagation distances where they achieve the maximum $a_0$. (c1)-(c3) The plasma density distribution at the central slice of the corresponding nonlinear wakes. The white dashed lines in (b1), (b3), (c1), (c3) represent the major axis of the laser transverse intensity profile and the $x$-axis.}
\end{figure}

A close investigation of the nonlinear plasma wake was conducted to understand the different injection dynamics when using lasers with different spatiotemporal distribution. Since a pre-ionized plasma is used in the simulations and there is no density downramp (except the plasma exit), the injection relies on the evolution of the laser pulse \cite{tsung2004near, mangles2004monoenergetic, geddes2004high, faure2004laser, kalmykov2009electron}. As the laser driver undergoes focusing \cite{xu2023generation} or diffraction \cite{PhysRevAccelBeams.26.091303}, it may induce an expansion of the wake, consequently reducing the velocity of the wake tail. When the electrons in the high-density sheath surrounding the nonlinear plasma wake move to the wake tail, they can be injected if they are faster than the wake tail \cite{xu2017high, xu2023generation}. Fig. \ref{fig:wake_velocity}(a) shows the evolution of $\frac{\mathrm{d}\xi_\mathrm{t}}{\mathrm{d}z}$ for three cases, where $\xi_\mathrm{t}$ is defined as the axial location of the tail of the first wake. The black dashed line in \ref{fig:wake_velocity}(a) is the etching velocity of the laser driver in the 3D nonlinear regime ($v_\mathrm{etch}\approx \frac{3}{2}\frac{\omega_\mathrm{p}^2}{\omega_0^2} c\approx 0.0052 c$) \cite{PhysRevSTAB.10.061301}. Due to the evolution of the laser spot size and the $a_0$, $\frac{\mathrm{d}\xi_\mathrm{t}}{\mathrm{d}z}$ deviates from $-\frac{v_\mathrm{etch}}{c}$. For case g, $\frac{\mathrm{d}\xi_\mathrm{t}}{\mathrm{d}z}$ reaches its minimum (-0.016) at $z\approx -1.1~\milli\meter$ where the laser reaches its maximum $a_0$, and induces an injection. The injection then ceases as a result of three synergistic effects: the decrease of $a_0$ [Fig. \ref{fig:laser_para_evo}(c)], the increase of $\frac{\mathrm{d}\xi_\mathrm{t}}{\mathrm{d}z}$ [Fig. \ref{fig:wake_velocity}(a)], and the beam loading effect from the $\sim30$ kA injected beam. At $z\approx -0.4~\milli\meter$, $\frac{\mathrm{d}\xi_\mathrm{t}}{\mathrm{d}z}$ decreases to $-0.013$ and remains relatively low for a long distance, resulting in the second injection from $z\approx-0.4~\milli\meter$ to $z\approx0.7~\milli\meter$. 

For the elliptical laser case and the realistic laser case, $\frac{\mathrm{d}\xi_\mathrm{t}}{\mathrm{d}z}$ reaches its minimum ($-0.011$) at $z\approx-1~\milli\meter$, slightly behind the location where the laser reaches its maximum $a_0$. However, the injection around this position is much weaker than the Gaussian laser case as shown in Fig. \ref{fig:laser_para_evo}(e). We show the transverse intensity distribution of the $\xi_{a_0}$-slice of the three laser pulses when they achieve the maximum $a_0$ in Fig. \ref{fig:wake_velocity}(b1)-(b3) and the corresponding plasma density distribution at the central slice of the nonlinear wake in Fig. \ref{fig:wake_velocity}(c1)-(c3). Here the central slice of the nonlinear wake is defined as the slice where the on-axis $E_z=0$. The Gaussian laser pulse has an axisymmetrical intensity profile [(b2)] and the sheath structure of its excited nonlinear wake is axisymmetric and sharp [(c2)]. The intensity profile of the elliptical laser is shown in Fig. \ref{fig:wake_velocity}(b3). {Due to the astigmatism, its major and minor axes have swapped relative to the initial state.} As shown in Fig. \ref{fig:wake_velocity}(c3), a sharp and high-density sheath only exists along the major axis of the elliptical laser while the sheath becomes wide and complicated along the minor axis. Correspondingly, the forward velocity of the sheath electrons when they reside at the wake tail depends on their azimuthal angles. The electrons at the angles of the laser major axis have similar forward velocities comparable to the Gaussian case while electrons at other angles exhibit smaller forward velocities \cite{PhysRevResearch.2.043319}, which reduces the injected charge. 

\begin{figure}
\centering
\includegraphics[width=1.0\linewidth]{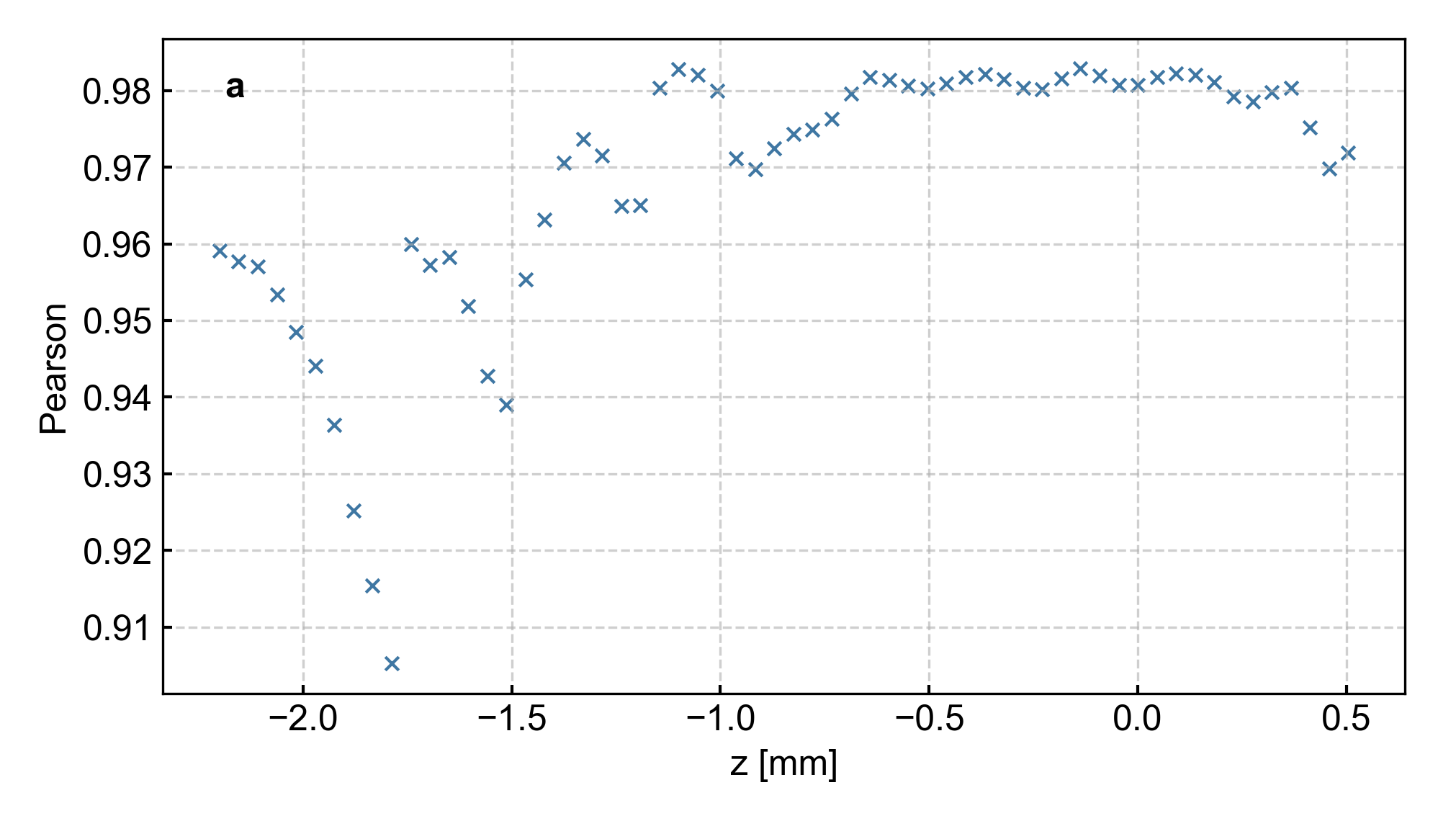}
\includegraphics[width=1.0\linewidth]{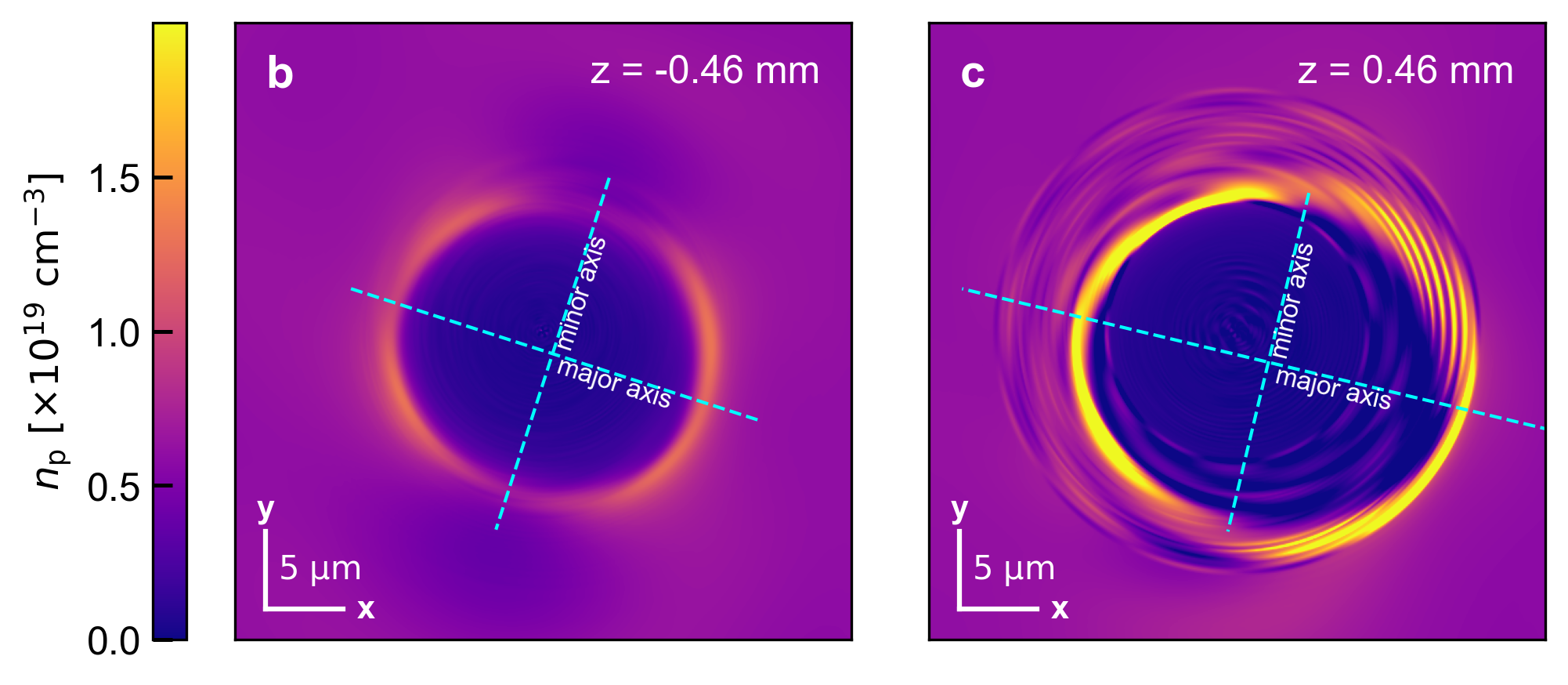}
\caption{\label{fig:two_wake} (a) The Pearson correlation coefficient between the realistic laser profile and its elliptical fit. The plasma density distribution at the central slice of the nonlinear wakes for the realistic laser pulse at $z=-0.46~\milli\meter$ (b) and $z=0.46~\milli\meter$ (c). The dashed lines represent the major and minor axes of laser driver.}
\end{figure}

The evolution of the realistic laser and its excited nonlinear wake are more complicated than the elliptical laser case. As shown in Fig. \ref{fig:wake_velocity}(c1), the overall sheath structure exhibits a larger width than that in the elliptical case. Especially, the sheath structure at the angles of the laser major axis is wide and diffuse. This significantly reduces the forward velocity of the sheath electrons \cite{Dalichaouch2021PoP} and suppress the injection at $z\approx -1 ~\milli\meter$ where $\frac{\mathrm{d}\xi_\mathrm{t}}{\mathrm{d}z}$ reaches a local minimum. As the realistic laser propagates inside the plasma, the complex spatial features beyond an ellipse dissipate gradually \cite{PhysRevLett.133.045002} and its transverse intensity distribution is closer to an ellipse, as shown in Fig. \ref{fig:two_wake}(a), {where the Pearson correlation coefficient is the covariance between the intensities of the realistic laser profile and the fitted elliptical profile, normalized by the product of their standard deviations \cite{enwiki:1335475625}. A value of the Pearson correlation coefficient close to 1 indicates strong agreement between the realistic and fitted profiles.} It is observed that at $z\approx-0.6~\milli\meter$, the realistic laser evolves into a well-defined elliptical laser, and propagates with an elliptical intensity profile in the plasma. Figs. \ref{fig:two_wake}(b) and (c) show the sheath structure at two different propagation distances. At $z=-0.46~\milli\meter$, the sheath width at the laser major axis is narrower than that at the minor axis. However, the wake radius at this distance is small and the phase velocity of the wake tail is close to $c$ [Fig. \ref{fig:wake_velocity}(a)], thus there is still no injection at this position. At $z=0.46~\milli\meter$, the wake radius grows, a narrow and high-density sheath is formed at the laser major axis, and the tail phase velocity is low. The combined effect of these three factors results in the significant injection as shown in Fig. \ref{fig:laser_para_evo}(e).  

Simulations indicate that the non-Gaussian transverse distribution of the laser pulse reduces the maximum self-focused intensity in an underdense plasma. The $a_0$ evolution for the realistic laser and the elliptical laser is similar, which suggests that the self-focusing of our laser pulses is mainly determined by the transversely elliptical intensity distribution. Meanwhile, due to the the lack of symmetric of the realistic laser intensity and phase profile, the sheath structure of its excited nonlinear plasma wake is wider and more complicated than that of the Gaussian laser case, which usually reduces the forward velocity of the sheath electrons and impedes the injection. As a result, the injected charge decreases progressively from the Gaussian laser case to the elliptical laser case to the realistic laser case, with the latter aligning with experimental results.

\section*{IV. Conclusion}
In conclusion, we conducted LWFA experiments aiming to produce relativistic electron beam with high charge. By applying the GS algorithm, we reconstructed the wavefront of the realistic laser pulse and imported the field distribution into the Q3D PIC simulations. By comparing the laser evolution, the nonlinear plasma wake excitation and the injection in three cases: a Gaussian laser pulse, an elliptical laser pulse and the reconstructed realistic laser pulse, we concluded that the realistic laser distribution hinder the injection through lowering the self-focused intensity and blurring the sheath structure of the nonlinear wake. As the realistic laser pulse propagates in the plasma, its intensity distribution becomes close to an ellipse which enables the injection. The simulation results of the injected charge when using the realistic laser pulse agree well with the experiments. This work indicates the importance of the optimization of the laser distribution when conducting LWFA experiments.

\section*{Acknowledgments}
\begin{acknowledgments}
We would like to thank Xuezhi Wu for helpful discussions. This work was supported by the Fundamental and Interdisciplinary Disciplines Breakthrough Plan of the Ministry of Education of China-JYB2025XDXM204, the National Grand Instrument Project No. SQ2019YFF01014400, Beijing outstanding young scientist project, National Natural Science Foundation of China No. 12375147, and the Fundamental Research Funds for the Central Universities, Peking University. The simulations were supported by the High-performance Computing Platform of Peking University and Tianhe new generation supercomputer at National Supercomputer Center in Tianjin.
\end{acknowledgments}

\section*{Data Availability}

The data that support the findings of this study are available from the corresponding author upon reasonable request.

\section*{Appendix A: Phase retrieval using the GS algorithm}
\begin{figure}
\centering
\includegraphics[width=0.75\linewidth]{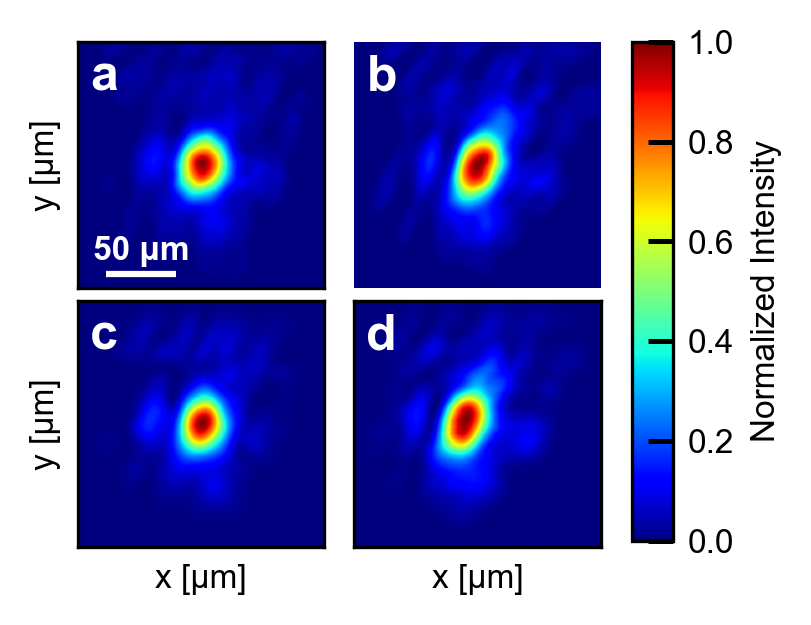}
\caption{GSA-MD results of the realistic laser pulse. (a) and (b) depict the measured intensity profiles at 0 and $-1~\milli\meter$, respectively. (c) and (d) are the retrieved intensity profiles at 0 and $-1~\milli\meter$, respectively. Note that a pixel in these plots represents a physical size of $0.44~\micro\meter \times 0.44~\micro\meter$. }
\label{figure:gsamd1}
\end{figure}

We measured the intensity profiles of the laser pulse using a Spiricon CCD SP620U, with an $\times10$ Apochromatic (APO) objective lens. Figs. \ref{figure:gsamd1}(a) and (b) show the measured intensity profiles at the focus and at a plane located 1 $\milli\meter$ upstream of the focus (labeled as `-1 $\milli\meter$'). By setting the intensity profiles as constraints, and initialize the phase distribution following Refs. \cite{Moulanier2023, Moulanier2024}, the phase distribution at the focal plane (far-field) was retrieved after several thousand iterations. The phase retrieval algorithm implemented in this study is based on the GSA-MD algorithm \cite{Moulanier2023, Moulanier2024}, which decomposes the electromagnetic (EM) fields into Hermite-Gaussian modes, thereby ensuring the correctness of laser propagation. The retrieved phase was validated through a back-propagation test. By using the retrieved far-field intensity [Fig. \ref{figure:gsamd1}(c)] and the retrieved phase, we calculated the intensity profile of a near-field plane ($z=-1~\milli\meter$) via angular spectrum (AS) diffraction code. Good agreement was achieved between the calculated near-field profile [Fig. \ref{figure:gsamd1}(d)] and the measured one [Fig. \ref{figure:gsamd1}(b)], which exhibits a structural similarity (SSIM) of 0.93 \cite{SSIM} and a peak signal-to-noise ratio (PSNR) of 37.5 dB in the image region \cite{PSNR}. Note that a higher SSIM (closer to 1) and a larger PSNR indicate better retrieved performance. 

To further validate the phase retrieval algorithm, we examined two additional benchmark cases, one using an ideal Gaussian laser pulse and the other using a 785 nm solid-state laser. For the Gaussian laser case, we chose $w_0=25.9~\micro\meter$ and numerically generated the intensity profiles at the focus and the $-2~\milli\meter$ plane. The near-field intensity profile calculated using the retrieved phase achieved exceptional fidelity with the goal, yielding a SSIM of 0.99999976 and a PSNR of 85.92 dB only after 2 iterations. The 785 nm laser coaxially propagated with the laser pulse driver used in the LWFA experiments and exhibited high-quality flat-top beam characteristics. We implemented the GSA-MD algorithm using intensity profiles in 3 planes: the focal plane, $-100~\micro\meter$, $-250~\micro\meter$. With an increase in the number of intensity profiles, the algorithm can skip local minima effectively. We compared the calculated intensity profile in the $-150~\micro\meter$ plane with the measured intensity and achieved a SSIM of 0.998 and a PSNR of 49.1 dB. {Following the guidelines in Ref. \cite{Moulanier2024}—namely using a high sampling density in the embedded Bayesian optimization and enforcing stringent convergence criteria—the phase retrieval algorithm produces consistent results across independent runs.}

\begin{figure}
\centering
\includegraphics[width=1.0\linewidth]{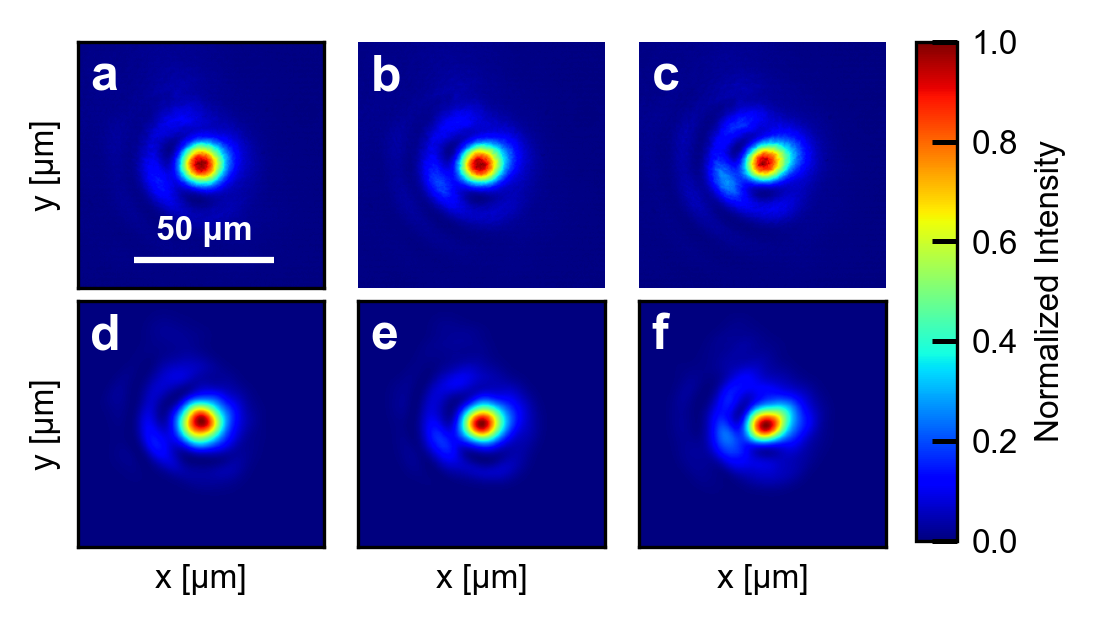}
\caption{GSA-MD results of a 785 nm solid-state laser.  (a), (b) and (c) depict the measured intensity profiles at 0, -100 and $-250~\micro\meter$, respectively. (d), (e) and (f) are the retrieved intensity distributions at 0, -100 and $-250~\micro\meter$, respectively. Note that a pixel represents a physical size of $0.44~\micro\meter \times 0.44~\micro\meter$.}
\label{figure:gsamd2}
\end{figure}

\begin{figure}
\centering
\includegraphics[width=1.0\linewidth]{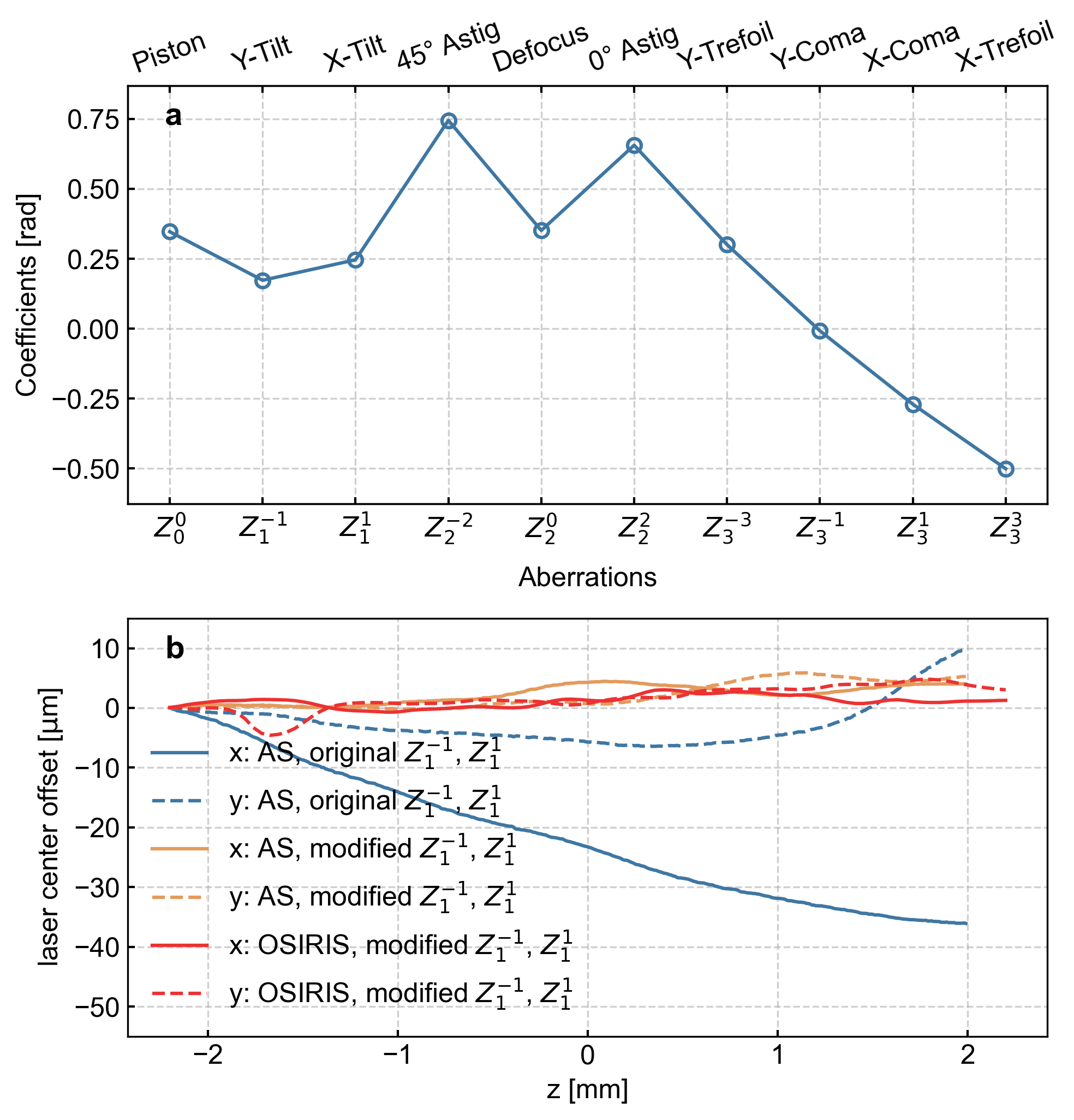}
\caption{ (a) The first 10 Zernike polynomials of the realistic laser phase. (b) The deflection of the transverse center of the laser in vacuum calculated using the AS diffraction code with original $Z_1^{-1}$ and $Z_1^1$ (blue lines), and with modified $Z_1^{-1}$ and $Z_1^1$ (orange lines). The red lines are the results in plasma obtained from OSIRIS.}
\label{figure:fig8}
\end{figure}

Zernike polynomials $Z_\mathrm{n}^\mathrm{m}(\rho,\theta)$ of the retrieved phase of the laser driver reveal the existence of low-amplitude, high-frequency noise, which is a known artifact of GS phase retrieval in low-intensity zones \cite{Fienup1982}. Thus, we implemented a Zernike-polynomials-based filter to eliminate $n>12$ polynomials. We then calculated the propagation of the laser in vauum using the AS diffraction code. As shown in Fig. \ref{figure:fig8}(a), we observed coefficients $Z_1^{-1}=0.173$ rad and $Z_1^{1}=0.246$ rad, corresponding to tilting components of the $y$ and $x$ directions, and a third-order aberration of $W_\mathrm{11}=\sqrt{(Z_1^{-1}-2Z_3^{-1})^2+(Z_1^{1}-2Z_3^{1})^2}=0.809$ rad, which can lead to an obvious deflection. Due to the inhomogeneous intensity distribution and the presence of other aberrations, the deflection of the laser center is complex and does not follow a simple linear trend, as illustrated by the blue lines in Fig. \ref{figure:fig8}(b). This large deviation in $x$-direction would significantly increase the PIC simulation cost due to the demand for a wider simulation box and more azimuthal modes. To avoid this, we modified these two tilting terms to let the laser have the minimum deflection. As shown by the orange lines in (b), the maximum offset in both directions is less than $5~\micro\meter$. The PIC simulations with a laser pulse with the modified terms also confirmed that the offset is small [red lines in (b)].

\section{APPENDIX B: 3D electromagnetic fields of the laser pulse}
The 3D spatial distribution of the laser's EM fields needs to be initialized at the entrance of the plasma in PIC simulations. Since the focal plane is at 1.2 mm inside the plasma, the initial laser pulse is $0.46z_\mathrm{R}$ away from the vacuum focus. We can use the AS diffraction code to give the complex amplitude of the laser central slice at the plasma entrance based on the reconstructed information at the focus. 

By assuming a Gaussian temporal envelope, the intensity and phase distribution at other transverse planes can be known as
\begin{align}
    I(z,x,y)&=I(z_\mathrm{n},x,y)\mathrm{exp}\left[ - 4\mathrm{ln}2\frac{(z-z_\mathrm{n})^2}{(c\tau_\mathrm{FWHM})^2}\right] \label{eq:I} \\
    \Phi (z,x,y) &= k(z-z_\mathrm{n}) + \Phi (z_\mathrm{n},x,y) \label{eq:Phi} 
\end{align}
where $z_\mathrm{n}$ is the center of the laser pulse, $I(z_\mathrm{n},x,y)$ and $\Phi (z_\mathrm{n},x,y)$ are the intensity and phase distributions measured or reconstructed in the experiments. Since the duration of the laser pulse is $\sim 30~\femto\second$, the radius of curvature of the wavefront and the difference of the Gouy phase over the ultrashort laser are not included in Eqs. \ref{eq:I} and \ref{eq:Phi}. Furthermore, since the laser pulse considered here contains several dozen cycles, the carrier-envelop phase doesn't play a critical role in the LWFA and it is chosen to let the peak of the envelope coincide with the peak of the fields. The EM fields of the laser pulse can be known straightforwardly from Eqs. \ref{eq:I} and \ref{eq:Phi}.

\begin{figure}
\centering
\includegraphics[width=1.0\linewidth]{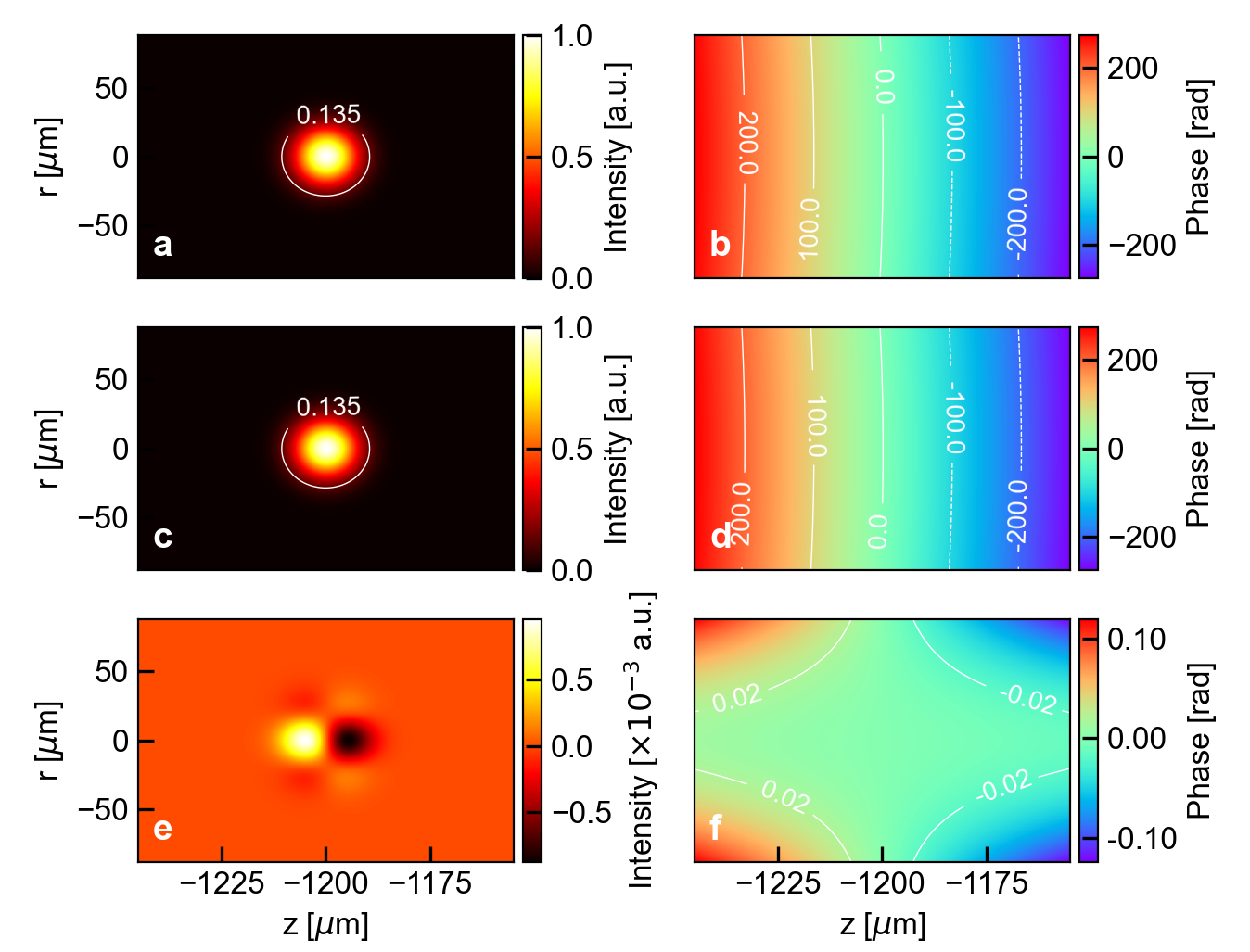}
\caption{\label{fig:fig9} Comparison between the formula [(a) for intensity and (b) for phase] and numerical results using Eqs. \ref{eq:I} and \ref{eq:Phi} [(c) for intensity and (d) for phase] for an ideal Gaussian laser. (e) is the intensity difference between (a) and (c), and (f) is the phase difference between (b) and (d).}
\end{figure}

The feasibility of building the 3D spatial distribution of the EM fields using the above assumption was demonstrated through the Gaussian laser pulse in case g. Fig. \ref{fig:fig9} (a) and (b) show the intensity and phase distributions calculated by the ideal Gaussian laser formula, while (c) and (d) display the corresponding distributions reconstructed using Eqs. \ref{eq:I} and \ref{eq:Phi} with the intensity and phase profiles at $z_\mathrm{n}=-1.2$ mm. As shown in Fig. \ref{fig:fig9} (e) and (f), the maximum intensity and phase differences are $9.6\times10^{-4}$ and 0.12 rad, respectively, both of which are negligible.

\section{APPENDIX C: Q3D OSIRIS SIMULATION CHECK}
\begin{figure}
\centering
\includegraphics[width=1.0\linewidth]{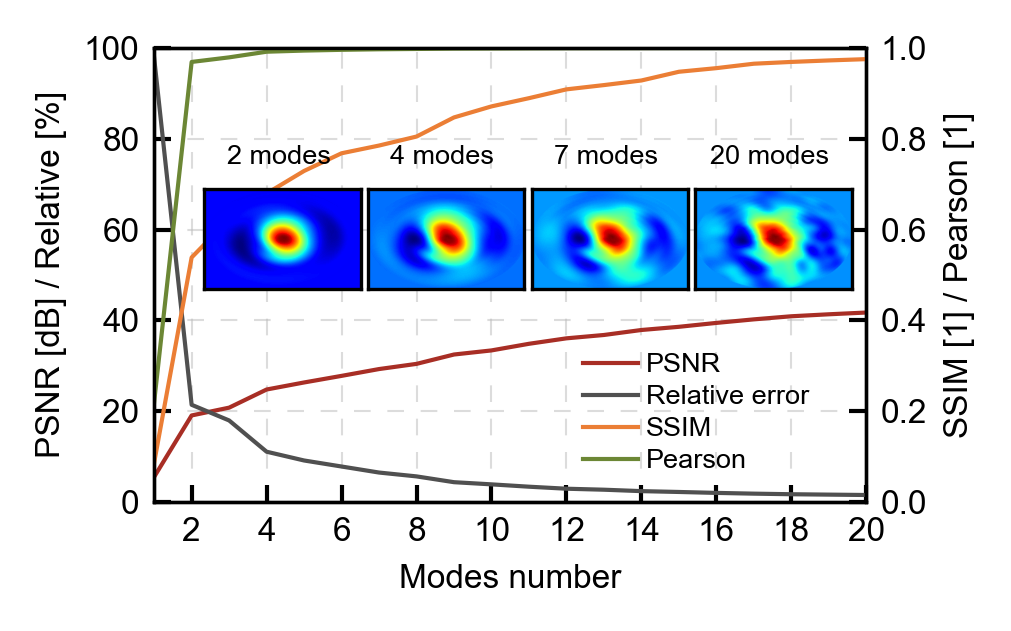}
\caption{\label{fig:error_line} The similarity between realistic laser spot and the laser spot used as OSIRIS input with different modes number. Gray line is relative error, orange line is SSIM, green line is Pearson correlation, and red line is PSNR. Insets are the reconstructed electric field slice with different modes number. }
\end{figure}

In the Q3D PIC code, all the physical quantities are decomposed into different azimuthal modes as $\mathrm{exp}(im\theta)$, where $m$ is an integer and $\theta$ is the azimuthal angle. Two modes ($m=0$ and $m=1$) are enough to model LWFAs driven by laser pulses with Gaussian transverse profiles. For realistic laser pulses with complicated transverse intensity and phase distributions, more modes are needed \cite{Zemzemi_2020}. The computational complexity of Q3D simulations scales as $(2n_\mathrm{modes}-1) \times n_\mathrm{z}\times n_\mathrm{r}$, where $n_\mathrm{modes}$ is the used modes number, $n_z$ and $n_r$ are the grid number along the $z$ and $r$ direction, respectively. Furthermore, the number of macro-particles per cell in the azimuthal direction is suggested to be $4(n_\mathrm{modes}-1)$ \cite{davidson2015implementation}. A large $n_\mathrm{modes}$ can model the realistic laser pulse more accurately but with a significantly increased computational cost. We firstly decomposed the initial 3D laser electric field as $\mathrm{exp}(im\theta)$ and then reconstructed the field using limited mode number. We compared the reconstructed field at the laser center with the original data and the results are shown in Fig. \ref{fig:error_line}. When $n_\mathrm{modes}=7$, it can accurately characterize the primary features of the laser spot with a PSNR exceeding 30 dB and a SSIM reaching 0.8. When $n_\mathrm{modes}=20$, high-fidelity results with a PSNR of 40 dB, a SSIM of 0.94, and a Pearson correlation coefficient of 0.99968 are achieved. After evaluating the trade-off between the computational cost and the accuracy, we chose $n_\mathrm{modes}=7$ in the simulations presented in the main body. 

\bibliographystyle{apsrev4-1}

\bibliography{refs_xinlu}
\end{document}